\documentclass[aps,prl,onecolumn,showpacs,superscriptaddress,groupedaddress,nofootinbib]{revtex4}  
\usepackage{color}
\usepackage{natbib}
\definecolor{light-gray}{gray}{0.45}
\definecolor{amber(sae/ece)}{rgb}{1.0, 0.49, 0.0}
\definecolor{ao(english)}{rgb}{0.0, 0.5, 0.0}
\definecolor{azure(colorwheel)}{rgb}{0.0, 0.5, 1.0}
\usepackage{amsmath,amsfonts,amssymb}
\usepackage{graphicx,array}
\usepackage[colorlinks=true, allcolors=blue]{hyperref}

\begin{document} 

\widetext
\title{Level anti-crossing magnetometry with color centers in diamond}

\author{Huijie Zheng}
\affiliation{Johannes Gutenberg-Universit{\"a}t  Mainz, 55128 Mainz, Germany}
%\email{zheng@uni-mainz.de}
\author{Georgios Chatzidrosos}
\affiliation{Johannes Gutenberg-Universit{\"a}t  Mainz, 55128 Mainz, Germany}
\author{Arne Wickenbrock}
\affiliation{Johannes Gutenberg-Universit{\"a}t  Mainz, 55128 Mainz, Germany}
\author{Lykourgos Bougas}
\affiliation{Johannes Gutenberg-Universit{\"a}t  Mainz, 55128 Mainz, Germany}
\author{Reinis Lazda}
\affiliation{Laser Centre, The University of Latvia, 19 Rainis Boulevard, LV-1586 Riga, Latvia}
\author{Andris Berzins}
\affiliation{Laser Centre, The University of Latvia, 19 Rainis Boulevard, LV-1586 Riga, Latvia}
\author{Florian Helmuth Gahbauer}
\affiliation{Laser Centre, The University of Latvia, 19 Rainis Boulevard, LV-1586 Riga, Latvia}
\author{Marcis Auzinsh}
\affiliation{Laser Centre, The University of Latvia, 19 Rainis Boulevard, LV-1586 Riga, Latvia}
\author{Ruvin Ferber}
\affiliation{Laser Centre, The University of Latvia, 19 Rainis Boulevard, LV-1586 Riga, Latvia}
\author{Dmitry Budker}
\affiliation{Johannes Gutenberg-Universit{\"a}t  Mainz, 55128 Mainz, Germany}
\affiliation{Helmholtz-Institut Mainz, 55128 Mainz, Germany}
\affiliation{Department of Physics, University of California, Berkeley, CA 94720-7300, USA}
\affiliation{Nuclear Science Division, Lawrence Berkeley National Laboratory, Berkeley, CA 94720, USA}
%\authorinfo{Further author information: Send correspondence to H.Z.: zheng@uni-mainz.de.}
% Option to view page numbers
\pagestyle{empty} % change to \pagestyle{plain} for page numbers   
\setcounter{page}{301} % Set start page numbering at e.g. 301

\begin{abstract}
Recent developments in magnetic field sensing with negatively charged nitrogen-vacancy centers (NV) in diamond employ magnetic-field (MF) dependent features in the photoluminescence (PL) and eliminate the need for microwaves (MW). Here, we study two approaches towards improving the magnetometric sensitivity using the ground-state level anti-crossing (GSLAC) feature of the NV center at a background MF of 102.4\,mT. Following the first approach, we investigate the feature parameters for precise alignment in a dilute diamond sample; the second approach extends the sensing protocol into absorption via detection of the GSLAC in the diamond transmission of a 1042\,nm laser beam. This leads to an increase of GSLAC contrast and results in a magnetometer with a sensitivity of 0.45\,nT/$\sqrt{\text{Hz}}$ and a photon shot-noise limited sensitivity of 12.2\,pT/$\sqrt{\rm{Hz}}$.
\end{abstract}
\maketitle

% Include a list of keywords after the abstract 
%\keywords{nitrogen-vacancy centers, level-crossing magnetometry, microwave-free magnetometry, absorption-based magnetometry}

\section{Introduction}
\label{sec:intro}  % \label{} allows reference to this section
\begin{figure}
\begin{center}
\begin{tabular}{c}
\includegraphics[width=\columnwidth]{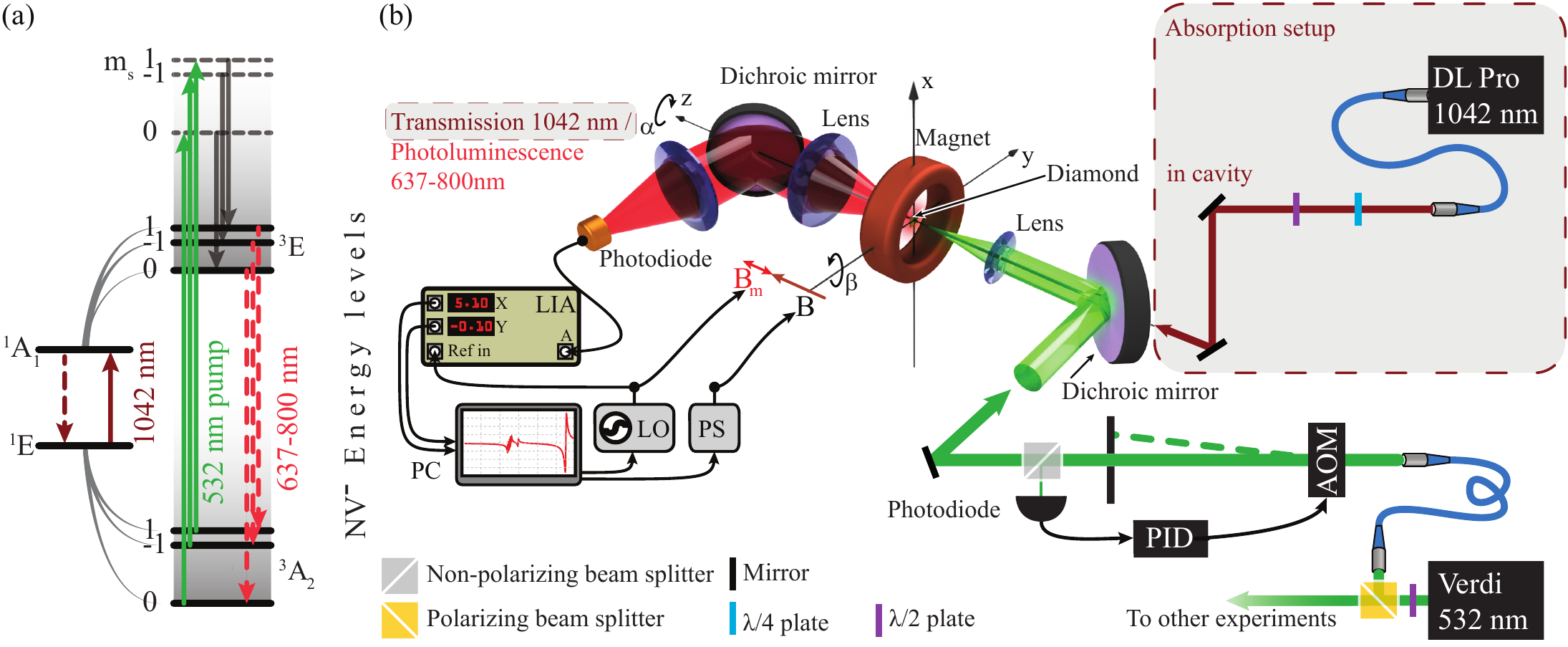}
\end{tabular}
\end{center}
\caption 
{ \label{Figure1} 
(a) NV-center energy level schematic. Solid green and red lines indicate excitations, dashed lines indicate radiative transitions, and gray solid lines indicate non-radiative transitions. (b) Schematic of the experimental setup. AOM: acousto-optic modulator, PID: proportional-integral-derivative controller, LIA: lock-in amplifier, LO: local oscillator, PS: power supply.} 
\end{figure}

Magnetic-field (MF) sensing is of importance in many applications in fundamental physics, biology, and materials science. Using the negatively-charged NV center in diamond is attractive due to its high magnetic sensitivity at ambient conditions for a given sensing volume\cite{Rondin2014,le2013optical,glenn2015single,balasubramanian2008nanoscale,maze2008nanoscale}. Diamond-based magnetic sensors are realized via measurements of the NV center's magnetically sensitive ground state, commonly by using optically detected magnetic resonance (ODMR) techniques\,\cite{ODMR1,Rondin2014,Wolf2015,SingleNVMagnetometry}. 
ODMR sensing protocols predominantly involve the use of green pump light for NV-center spin polarization, application of MW fields for spin manipulation, and an optical readout step involving either detection of NV-PL or absorption on the singlet transition at 1042\,nm [Fig.~\ref{Figure1}\,(a)].
However, there are applications, e.g., nano-magnetic resonance imaging \cite{ajoy2016dc}, eddy current detection \cite{MIT2,MIT3a} and MF mapping of conductive, magnetic structures\cite{Simpson2016} where the use of strong, continuous wave (cw) or pulsed, MW fields employed in MW-based ODMR is intolerable.\\
\indent Recently, we demonstrated a novel MW-free magnetometric protocol based on the properties of the NV-center's ground-state level anti-crossing (GSLAC) \cite{wickenbrock2016microwave}. Applying a $\sim$102.4\,mT background MF causes an avoided crossing between two of the ground-state NV Zeeman-sublevels, resulting in spin-population transfer observable in changes of the NV-PL or 1042\,nm absorption. The resulting MF-dependent feature can be used for sensitive magnetometry \cite{wickenbrock2016microwave,ajoy2016dc,Wood2016arxiv}.\\
\indent In this work, we explore two distinct avenues towards improved magnetometric sensitivities using the MW-free sensing protocol. 
We investigate the GSLAC lineshape as a function of nitrogen concentration, [N], and MF alignment, and, for the first time, we implement a magnetometer based on the GSLAC feature in absorption on the singlet transition $^{1}$E $\rightarrow$ $^{3}$A$_{2}$ [Fig.\,\ref{Figure1}(a)]. Additionally, we compare our recently published results\,\cite{wickenbrock2016microwave} to a comparable diamond in a highly homogeneous magnet in Riga, to exclude MF gradient-related broadening.\\% with a demonstrated sensitivity of 0.45\,nT/$\sqrt{\rm{Hz}}$.

\section{Experimental setups}

The experiments where conducted on three different setups, two in Mainz and one in Riga.\\
\indent A combined schematic of the experimental setups in Mainz is shown in Fig.~\ref{Figure1}(b); the first one is fluorescence-based and allows us to perform measurements on different samples with complete and precise control over all degrees of freedom in alignment, while the other one is an absorption-based magnetometer. In both setups, the NV centers in the diamond samples are optically spin-polarized with power-stabilized 532\,nm light provided by a diode-pumped solid-state laser (Coherent Verdi V10). Details of the optical and electrical components in the fluorescence-detection setup can be found in Ref.\, \cite{wickenbrock2016microwave}. In the absorption-detection setup, 1042\,nm light used to probe the singlet transition is delivered by a fiber-coupled extended-cavity diode laser (Toptica DL Pro) and locked to an optical cavity, which consists of the diamond sample with appropriate coating on either side and a spherical mirror. The absorption-magnetometry method and the setup are based on an improved version of a recently demonstrated cavity-enhanced NV magnetometer\cite{NVCavity1} and a more detailed description of the current experimental improvements is presented in Ref.\,\cite{gechatzi}.\\
\indent The diamond samples in both setups are placed within custom-made electromagnets of the same build. They have 200\,turns in a 1.3\,cm thick coil with a 5\,cm bore, are wound on a water-cooled copper mount, and produce a background field, B, of 2.9\,mT per ampere supplied. For a field of 120\,mT, approximately 1.8\,kW are dissipated. The current is provided by a computer-controlled power supply (Keysight N8737A).\\
\indent In the fluorescence-detection setup the diamond can be rotated also around the z-axis [Fig.~\ref{Figure1}(b); angle $\alpha$]. Moreover, the electromagnet can be moved with a computer-controlled 3D translation stage (Thorlabs PT3-Z8) and a rotation stage (Thorlabs NR360S, x-axis) [Fig.~\ref{Figure1}(b); angle $\beta$]. Therefore, in this setup, all degrees of freedom for placing the diamond in the center of the magnet and aligning the NV axis parallel to the MF can be addressed with high precision.\\
\indent In the absorption-based setup, the electromagnet is mounted on a manual 3D translation stage. An additional secondary coil (15 turns, gauge 22 wire, inner diameter of 12.5\,mm) is used to apply a small MF modulation, $\text{B}_m$, to the background field that allows for phase sensitive detection for magnetometric measurements. Its current is supplied by a function generator (Tektronix AFG2021),  which acts also as the local oscillator (LO) for a lock-in amplifier (LIA; SRS 865).\\
\indent The fluorescence-detection setup in Riga employed a custom-built magnet initially designed for electron paramagnetic resonance (EPR) experiments. It consists of two 19\,cm diameter iron poles with a length of 13\,cm each, separated by a 5.5\,cm air gap. This magnet could provide a highly homogeneous field. The diamond sample under investigation is held in place using a non-magnetic holder, allowing also for alignment of the NV axis to the applied MF. Green 532\,nm light (Coherent Verdi) is delivered to the sample via 400\,$\mu$m diameter core optical fiber (numerical aperture of 0.39). The same fiber is used for PL collection, which is separated from the residual green reflections by a long-pass filter (Thorlabs FEL0600) and focused onto an amplified photodiode (Thorlabs PDA36A-EC). The signals are recorded and averaged on a digital oscilloscope (Agilent DSO5014A).\\  %The resulting power is 115\,mW. 
\indent Finally, in Table\,\ref{table1} we present the characteristics of all the diamond samples used in this work. We note here, that the measurements using diamond sample F11 were previously reported in Ref.\,\cite{wickenbrock2016microwave}. The sample was initially built into the PL-detection setup in Mainz, and for this work we replaced it with a more dilute sample W4. In addition, for the PL-detection measurements performed in Riga, we used sample C7. In the absorption-detection setup we used the sample B3A, which has dielectric coatings on both sides. In particular, one side of the B3A diamond has a highly reflective (98.5\%) coating while the other side has an anti-reflective coating for 1042\,nm. This way, with an additional spherical mirror, the optical cavity to enhance the absorption on the singlet transition is formed\,\cite{gechatzi}.\\
\begin{figure}
  \centering
  \includegraphics[width=\columnwidth]{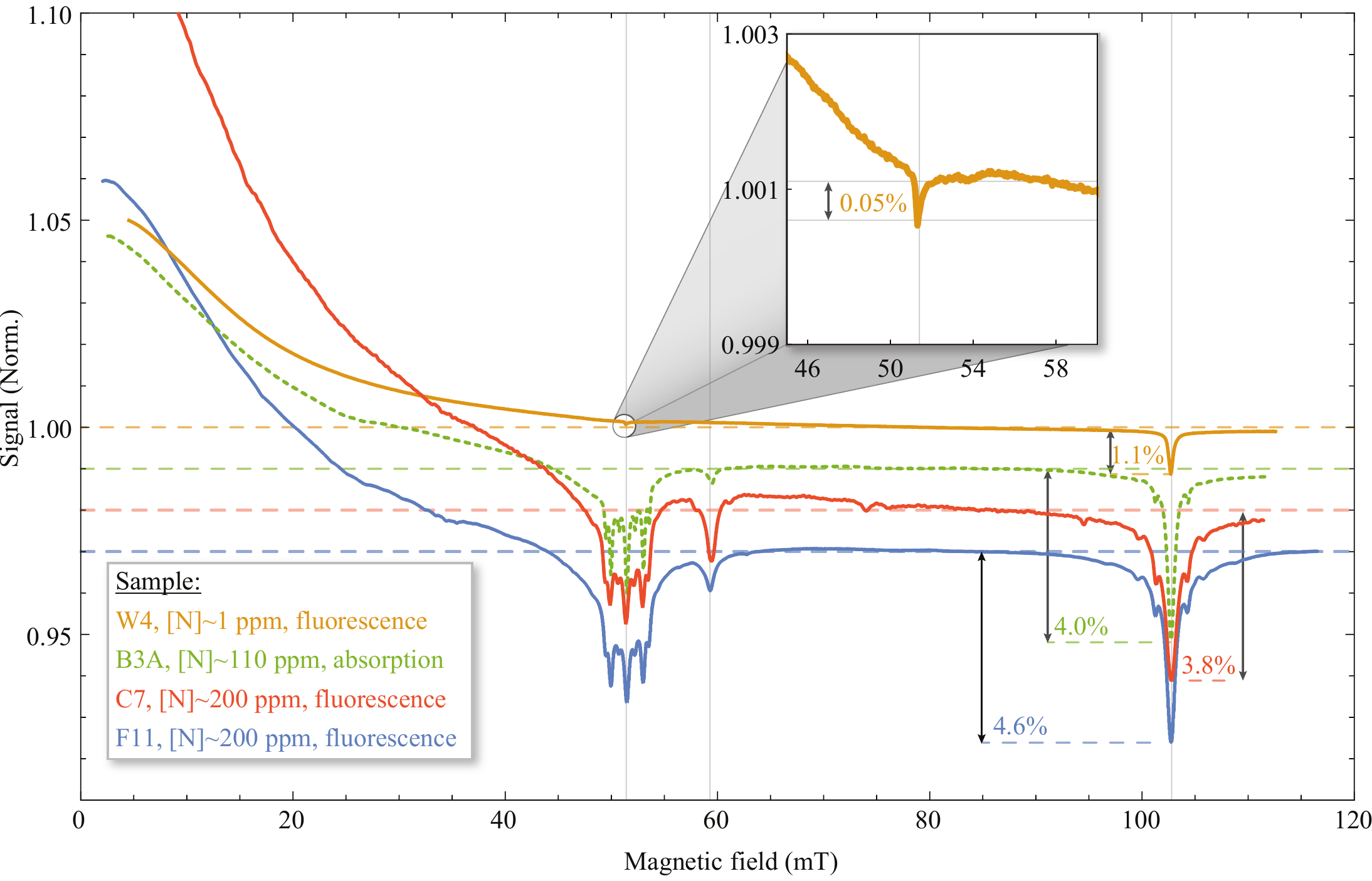}
  \caption{Traces of photoluminescence (solid) and absorption (dotted) signals for different [N]-density diamond samples as a function of the applied MF, normalized to their respective signals at 80\,mT. For better visibility the traces have been offset as is indicated by the dotted lines. The observed contrasts are shown explicitly for the GSLAC feature for the different traces. The inset shows a detailed view on the W4 trace around 51.2\,mT.}\label{Figure2}
\end{figure}

\begin{table*}[ht]
  \centering
\begin{tabular}{|m{5cm}||m{2.3cm}|m{2.3cm}|m{2.3cm}|m{2.3cm}|}
 \hline
Sample &\textbf{W4} &\textbf{B3A}&\textbf{F11} & \textbf{C7}\\
 \hline
 \hline
Type & CVD & HPHT & HPHT & HPHT\\
 \hline
Surface cut & (100) & (111) & (111) & (100)\\
\hline
[N] (ppm) & $<$ 1  & $<$ 110  &  $<$ 200 & $<$ 200 \\\hline
e$^-$ irradiation dosage (cm$^{-2}$)& 10$^{18}$  & 2$\times$10$^{19}$   & 10$^{18}$ & 10$^{18}$\\
\hline
e$^-$ irradiation energy (MeV)& 10  & 10  & 10& 10\\
\hline
Sample annealing &720\,$^o$C, 2\,h&700\,$^o$C, 2\,h&700\,$^o$C, 3\,h&750\,$^o$C, 3\,h\\
\hline
\end{tabular}
\caption{Diamond samples. CVD: chemical vapor deposition, HPHT: high pressure, high temperature.}
\label{table1}
\end{table*}

\section{Results and discussion}
Here we present a systematic investigation of relevant parameters in level anti-crossing magnetometry with NV centers towards highly sensitive MF measurements. For a level anti-crossing based magnetometric protocol, the attainable photon-shot-noise-limited MF sensitivity is proportional to\,\cite{NVCavity1}:
\begin{eqnarray}
\delta B (\rm{T}/\sqrt{\rm{Hz}})&\approx&\frac{1}{^{\gamma}/_{2\pi}}\,\frac{\Delta \nu_{\rm{mr}}}{C\sqrt{\mathcal{R}}},\label{eq:eq1}
\end{eqnarray}
where $|^{\gamma}/_{2\pi}| \simeq 28.024\,\rm{GHz}\,\rm{T}^{-1}$ is the gyromagnetic ratio of the electron spin, and $\mathcal{R}$ is the rate of detected photons in either PL or absorption measurements. $\Delta \nu_{\rm{mr}}$ and $C$ are the full width at half maximum (FWHM) linewidth and contrast of the GSLAC feature, respectively. It follows that, for a given photon-collection rate $\mathcal{R}$, to achieve the highest MF sensitivity, the ratio of contrast to linewidth needs to be maximized.\\

%++++++++++++++++ F I G U R E   2 ++++++++++++++++++++++++++++++++

\indent In Fig.\,\ref{Figure2} we present normalized PL and absorption measurements as a function of the background MF following an initial alignment of the electromagnet. This figure gives an overview of the changes in contrast and linewidth of the observable, anti-crossing, features for all the samples listed in Table\,\ref{table1}. The MF for W4 and B3A is scanned from 0 to 110\,mT in 10\,s, and the presented signal is the average of 64 traces. The MF of the EPR magnet in Riga is scanned in 100\,s from 0 to 120\,mT, and the presented signal is the average of 35 traces. The PL data using sample F11 are taken from Ref.\,\cite{wickenbrock2016microwave}. The presented traces contain several features extensively discussed in past\cite{Armstrong1,Hall2016,PhysRevB.94.155402} and more recent\cite{wickenbrock2016microwave, Ivanov2016, Wood2016_2} works. In particular,
the initial gradual decrease in the observed signals is associated either with a reduction in PL emission (samples W4, C7, F11; Fig.\,\ref{Figure1}), or with an increase in absorption (sample B3A; Fig.\,\ref{Figure1}) from the non-aligned NV centers due to spin-mixing. When a magnetic field is applied not along the NV-axis, it mixes the Zeeman sublevels. This resulting spin mixing reduces the effect of the optical pumping, and thus, decreases the population of the $^3$A$_2$ m$_s$=0 spin state and increases the population of the metastable singlet state. Moreover, around 51.2\,mT, the observed features for samples F11, B3A and C7 correspond to cross-relaxation between the NV center and substitutional nitrogen (P1) centers. We note here, that, for the most dilute sample used in this work (W4) we observe a significantly different structure. At a field of 51.2(1)\,mT (calibrated with microwave-spectroscopy measurements; not shown here) we observe a small drop in PL (contrast 0.05\%) that could be attributed to the excited state level-anticrossing (ESLAC) of the NV center. Detailed investigation of its origin will be the subject of future work. The feature at 60\,mT is attributed to cross relaxation with NV centers that are not aligned along the MF\cite{Armstrong1,wickenbrock2016microwave,Ivanov2016}. At $\sim$102.4\,mT we observe the feature attributed to the GSLAC of the NV center. Several additional features are visible, however, here, we focus on the contrast and linewidth of the central component due to their relevance to magnetometry applications (see Eq.\ref{eq:eq1}).
Finally, we tried to rule out MF gradients as the limitation of the GSLAC-feature width as reported before for sample F11\cite{wickenbrock2016microwave}. Therefore, a sample C7 with a comparable NV density was investigated in a highly homogeneous EPR magnet in Riga. However, due to alignment constraints, the results were inconclusive. We add them here for completeness.

%++++++++++++++++ F I G U R E   3 ++++++++++++++++++++++++++++++++
\begin{figure}[h!]
  \centering
  \includegraphics[width=\columnwidth]{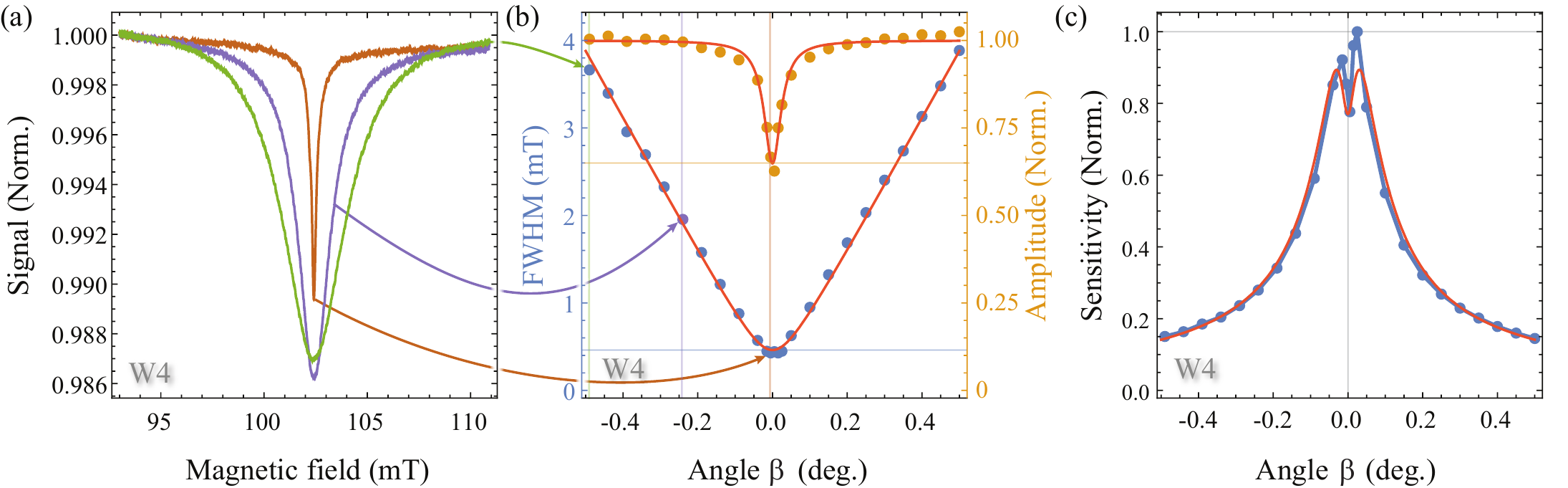}
  \caption{(a) GSLAC fluorescence contrast of diamond sample W4 as a function of the applied MF at different y-axis angles to the NV axis. (b) GSLAC FWHM width (blue dots) and contrast amplitude (amber dots) as a function of misalignment angle (x-axis). The experimental data are represented with dots, and the fits with solid lines. (c) Magnetic-field sensitivity (amplitude/width) as a function of misalignment angle $\beta$, normalized to the maximum obtained sensitivity.
}\label{Figure3}
\end{figure}

In Fig.\,\ref{Figure3} we present PL-based measurements investigating the dependence of the GSLAC-feature lineshape on MF alignment for the dilute sample W4, which displays the narrowest linewidth, as seen in Fig.\,\ref{Figure2}. In particular, the angles $\alpha$ (z-axis) and $\beta$ (y-axis) between the NV-axis and the applied MF are controlled with a precision better than 0.01\,degrees. Following an initial alignment optimization of both angles towards minimal GSLAC linewidth, we record traces of the feature as a function of the angle $\beta$. Figure.\,\ref{Figure3}\,(a) shows three examples of the recorded traces for different values of $\beta$. The data are then fitted with a Lorentzian function to extract the GSLAC feature's FWHM linewidth and contrast amplitude. The resulting data are displayed in Fig.\,\ref{Figure3}\,(b). While decreasing $\beta$ from large misalignment angles ($\left|\beta\right|>0.1$\,degrees), the linewidth reduces linearly towards a minimal value of $\Delta \nu_{\rm{mr}}=0.46(2)$\,mT. The contrast amplitude, however, stays constant and then sharply decreases to less than 65\% of its value for large misalignment. The contrast-amplitude data are presented in Fig.\,\ref{Figure3}\,(b), and are fitted with a Lorentzian function, yielding a FWHM width for the observed feature of 54(4)$\times 10^{-3}$\,degrees. This observed feature, resulting from a misalignment angle between the MF and the NV-axis, can be translated into a transverse MF-component of 97(7)\,$\mu$T in magnitude. This behavior can be interesting for applications of transverse MF sensing (similarly to the work presented in Ref.\,\cite{Wood2016arxiv}), which we will pursue further in the future. For longitudinal MF sensing however, this effect leads to an undesirable behavior, which we demonstrate in Fig.\,\ref{Figure3}\,(c). Here we present as a measure for longitudinal MF sensitivity (Eq.\,\ref{eq:eq1}), the ratio of contrast to linewidth. Due to the different angle dependence of the GSLAC feature width and the contrast amplitude, it appears that the sensitivity has a local minimum for optimum angle alignment ($\beta=0$) and a maximum for a non-zero angle of $\left|\beta\right|=0.01$\,degrees. At this angle, the amplitude of the GSLAC feature is highly sensitive to transverse magnetic fields, as well as to mechanical angle fluctuations which will appear as an additional (non-magnetic) noise source.
%in the sensing protocol. This makes a linewidth reduction via sample dilution unsuitable to improve the magnetometric sensitivity (at least to longitudinal magnetic fields).\\%%, the resulting FWHM from the Lorentzian function fit to the amplitude in Fig.\,\ref{Figure3} corresponds to a transverse magnetic field of )
%the width is decreasing gradually while the amplitude experiences a sharp reduction after a critical misaligned angle,
%The amplitude feature is much narrower than the width feature and could potentially be utilized as a transverse-field magnetometer. Taking into account the characteristics of both width and amplitude; the maximum sensitivity is achieved not when the MF and the NV-axis are aligned perfectly but when a small misalignment exists between the two.
%Combining the characteristics of the width and amplitude, the sensitivity is highest at a small angle tilted between NV-axis and magnetic field direction instead of them exactly aligned. This may result from a combination of coupling strength and decoherence. \color{azure(colorwheel)} (we can say something like "Taking into account the characteristics of both width and amplitude; the maximum sensitivity is achieved not when the MF and the NV-axis are aligned perfectly but when a small misalignment exists between the two.") \color{black} \\

\begin{figure}[h!]
  \centering
  \includegraphics[width=\columnwidth]{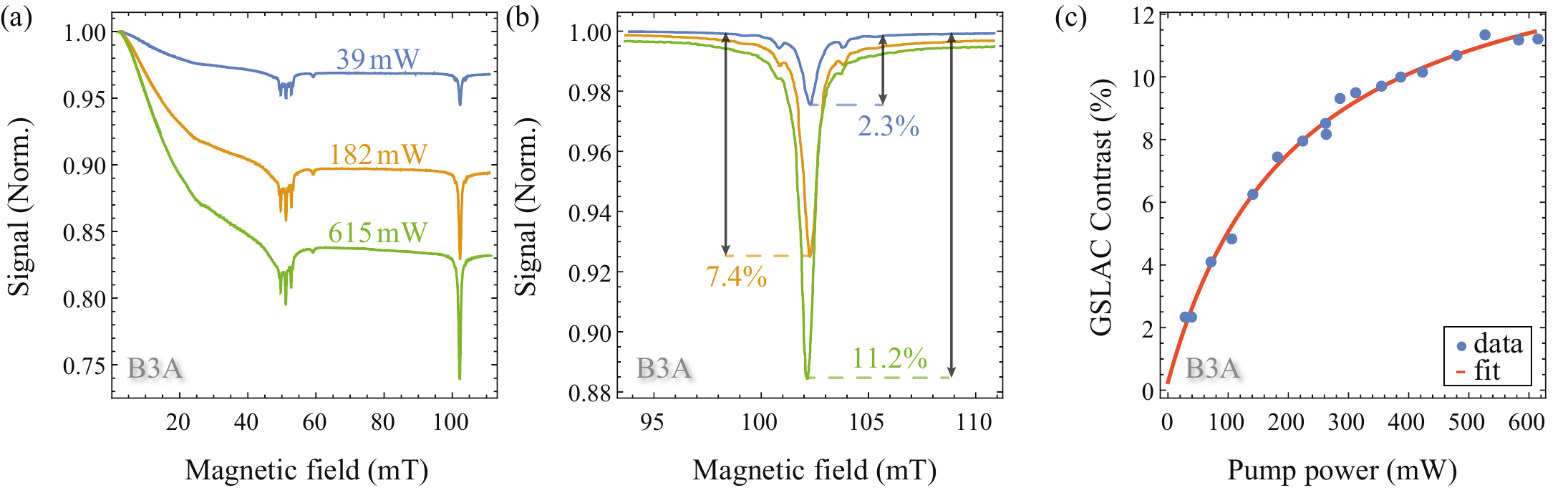}
  \caption{(a) Transmission signal of 1042\,nm light through the optical cavity utilizing sample B3A, as a function of applied magnetic field, and normalized to the measured transmission at zero-field. The three traces presented correspond to different pump-light powers. (b) Detail of the magnetic-field scan around the GSLAC feature for different pump-light powers normalized to the background transmission at 80\,mT. (c) GSLAC-feature contrast amplitude as a function of pump-light power.
  }\label{Figure4}
\end{figure}

%++++++++++++++++ F I G U R E   4 ++++++++++++++++++++++++++++++++
\indent In Fig.\,\ref{Figure4} we present transmission measurements of 1042\,nm light propagating through a cavity-enhanced absorption-based magnetometer utilizing the B3A sample, as a function of the applied background MF field and of the 532\,nm pump-light power. We note here, that for the measurements presented in Fig.\,\ref{Figure4}\,(a),\,(b),\,\&\,(c), the 532\,nm light-beam spot-size and 1042\,nm light-beam spot-size on the diamond were similar and approximately equal to $\sim$50\,$\mu$m. In particular, in Fig.\,\ref{Figure4}\,(a) \& (b) we present three examples of the recorded traces for three different values of 532\,nm light power. While we observe a similar behavior and features as for the PL-detection measurements (see Fig.\,\ref{Figure2} and the preceding discussion), a different initial signal drop, as well as, different GSLAC contrast amplitudes are observed for different 532\,nm light powers. Figure.\ref{Figure4}\,(b) shows a detailed expansion of the GSLAC-feature for the three different 532\,nm light powers used in Fig.\,\ref{Figure4}\,(a), along with the respective contrast amplitude. Moreover, a shift in the position of the feature, caused by a temperature increase due to the 532\,nm pump light is observed (see, for example, Ref\,\cite{VMA2010} for a discussion of the temperature dependence of the ground state $^3$A$_2$ splitting). All recorded traces at different 532\,nm light powers are fitted with a Lorentzian function, allowing us to extract the GSLAC feature's FWHM linewidth and contrast amplitude. The resulting data for the GSLAC contrast amplitude are displayed in Fig.\,\ref{Figure4}\,(c), showing a saturating behavior that yields a maximum attainable contrast of 15\% [resulting from the fit shown in Fig.\,\ref{Figure4}\,(c)]. We did not observe a significant change in the GSLAC FWHM as a function of the 532\,nm light power, and the average GSLAC FWHM of the recorded data is 0.84(1)\,mT. \\

% The data are fitted with a saturation curve $A-\frac{\textbf{B}}{C+x}$ where x represents pump light power and A the contrast at saturation. The maximum contrast from this fit is calculated to be 15\%.\\ 
%\color{azure(colorwheel)} or "The best fit parameter for A was 15 and values of 3$\times 10^{3}$ mW and 215 mW  were obtained for B and C respectively, indicating a contrast of $\sim$ 15\% at a few tens of W of pump light power".\color{black}

%++++++++++++++++ F I G U R E   5 ++++++++++++++++++++++++++++++++
\begin{figure}[b!]
  \centering
  \includegraphics[width=\columnwidth]{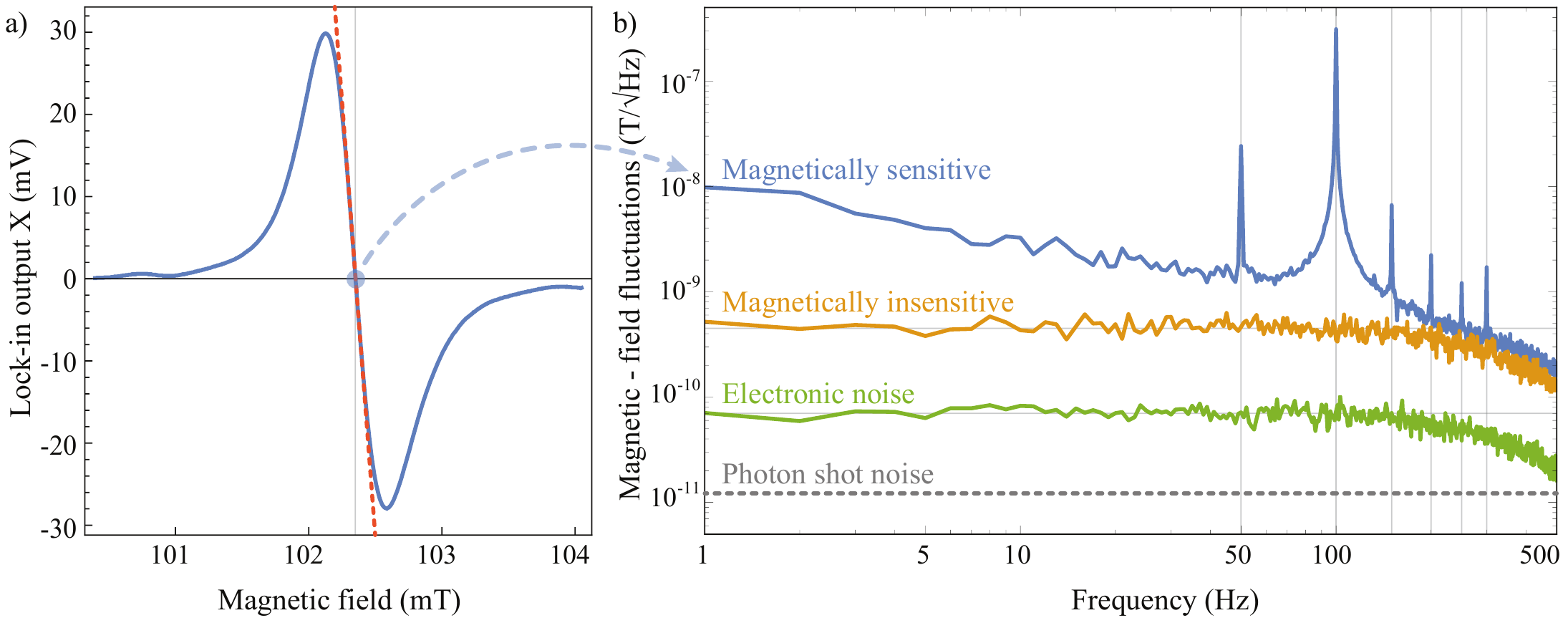}
  \caption{Absorption-based magnetometry with sample B3A. (a) Detail of the LIA output around the GSLAC feature (blue) with a linear fit to the data (red, dotted). The fit is used to calibrate the magnitude of the magnetic field fluctuations. (b) Noise of the magnetometer: magnetically sensitive at a field of 102.4\,mT, magnetically insensitive at a field of 80\,mT (average noise between 1$-$100\,Hz is  0.45\,nT/$\sqrt{\rm{Hz}}$), and electronic noise with no cavity transmission (average noise between 1$-$100\,Hz is 70\,pT/$\sqrt{\rm{Hz}}$). The photon shot noise limit of the magnetometer is indicated at 12.2\,pT/$\sqrt{\rm{Hz}}$. The decrease in signal for frequencies above 200\,Hz is due to the filtering of the LIA.}
\label{Figure5}
\end{figure}
\indent Finally, we demonstrate the MF sensitivity of the implemented absorption-based magnetometer utilizing sample B3A, using the GSLAC feature. For a 532\,nm pump-light power of $\sim\,600$\,mW, we record the transmission of 1042\,nm light locked on resonance to the optical cavity, while bringing the NV-center's energy levels into the GSLAC (102.4\,mT). In particular, by scanning the background MF around the
GSLAC feature while applying a small oscillating MF (B$_m\simeq\,0.01$\,mT at the modulation  frequency of 15\,kHz), we record the transmission signal using a photodiode (Thorlabs, PDA36A-EC). We obtain the signal-component oscillating at the modulation frequency using a LIA (SRS\,865; demodulation time constant 3\,ms). The resulting demodulated absorption signal (as detected in the properly phased LIA X output) is presented in Fig.\,\ref{Figure5}\,(a). It depends linearly on the applied background MF around the GSLAC, and can, therefore, be used for precise magnetometric measurements. Thus, by setting the background MF field value exactly to the center of the GSLAC feature (102.4\,mT) we record the transmission signal for an acquisition time of 1\,s. In Fig.\,\ref{Figure5}\,(b) we present the resulting MF noise-spectrum of the acquired data. We observe a 1/f MF sensitivity limited by the noise of the electromagnet-current power supply and ambient noise, and demonstrate a noise floor of magnetically insensitive measurements corresponding to 0.45\,nT/$\sqrt{\rm{Hz}}$. The peaks at 50\,Hz and harmonics are attributed to magnetic noise in the lab and are not visible on the magnetically insensitive spectrum, which is obtained operating at a background MF value of 80\,mT. The electronic noise floor was measured as 70\,pT/$\sqrt{\rm{Hz}}$. The photon-shot-noise limit is calculated to be 12.2\,pT/$\sqrt{\rm{Hz}}$ for 4.2\,mW of collected 1042\,nm light (Eq.\,\ref{eq:eq1}), and the quantum-projection-noise limit, related to the number of NV centers we probe, is calculated to be 0.7\,pT/$\sqrt{\rm{Hz}}$. 

\section{Conclusions}
In this article, we investigate two approaches to increasing the magnetometric sensitivity in microwave-free diamond-based magnetometers using the GSLAC of the NV center.
Sensitivity gains via feature width-reduction are problematic due to an experimentally observed amplitude decrease for a dilute sample at optimum alignment. The measured feature in misalignment angle is very narrow [corresponding to a transverse MF of 97(7)\,$\mu$T] and has the potential to be used for transverse MF magnetometry. For magnetometry along the background MF, however, a more promising route is the increase of signal amplitude in an absorption-based setup.
We demonstrate an improved microwave-free magnetometer setup based on a cavity-enhanced singlet-absorption GSLAC measurement, which exhibits an average noise floor of 0.45\,nT/$\sqrt{\rm{Hz}}$, and for our experimental conditions a photon-shot-noise limit of 12.2\,pT/$\sqrt{\rm{Hz}}$.\\
\indent Future investigations will involve a thorough study of the lineshape and width of the signal near the GSLAC and ESLAC, as well as the additional features around it, with the aim of understanding the fundamental sensitivity and bandwidth limitations\,\cite{BandwidthResponse} of our sensing protocol.\\

\acknowledgments % equivalent to \section*{ACKNOWLEDGMENTS}   
We acknowledge support by the DFG through the DIP program (FO 703/2-1). HZ is a recipient of a fellowship through GRK Symmetry Breaking (DFG/GRK 1581). GC acknowledges support by the internal funding of JGU. The Riga group gratefully acknowledges financial support from the Latvian National Research Programme (VPP) project IMIS2. LB is supported by a Marie Curie Individual Fellowship within the second Horizon 2020 Work Programme. DB acknowledges support from the AFOSR/DARPA QuASAR program. We thank Victor\,M.~Acosta and J.\,W.~Blanchard for fruitful discussions.

\bibliographystyle{apsrev}
\bibliography{report.bib} 
%\bibliography{report.bib} % bibliography data in report.bib
%\bibliographystyle{spiebib} % makes bibtex use spiebib.bst

\end{document}